# Weak charge-lattice coupling requires reinterpretation of stripes of charge order in La$_{1-x}$Ca$_x$MnO$_3$

J.C. Loudon[1], S. Cox[1], A. J. Williams[2], J. P. Attfield[2], P. B. Littlewood[3], P. A. Midgley[1]

and N. D. Mathur[1a]

[1]*Department of Materials Science, University of Cambridge, Pembroke Street, Cambridge CB2 3QZ, UK*

[2]*School of Chemistry and Centre for Science at Extreme Conditions, University of Edinburgh, West Mains Road, Edinburgh EH9 3JJ, UK*

[3]*Cavendish Laboratory, University of Cambridge, Madingley Road, Cambridge, CB3 0HE, UK*

Modulations in manganites attributed to stripes of charge/orbital/spin order are thought to result from strong electron-lattice interactions that lock the superlattice and parent lattice periodicities. Surprisingly in La$_{1-x}$Ca$_x$MnO$_3$ (*x*>0.5, 90 K), convergent beam (3.6 nm spot) electron diffraction patterns rule out charge stacking faults and indicate a superlattice with uniform periodicity. Moreover, large area electron diffraction peaks are sharper than simulations with stacking faults. Since the electron-lattice coupling does not lock the two periodicities (to yield stripes) it may be too weak to strongly localise charge.

71.38.-k, 71.45.Lr, 73.20.Mf, 75.47.Lx

[a] Corresponding author: ndm12@cus.cam.ac.uk

Stripe and chequerboard phases have been understood to arise in complex oxides such as manganites[1-11], cuprates[12], nickelates[13], cobaltites[14] and ferrites[15]. The pseudo-cubic perovskite manganites ($RE^{3+}_{1-x}AE^{2+}_xMnO_3$, RE=rare earth, AE=alkaline earth) are characterised by long-range modulations with wavevector **q**∥**a***, indexing the room temperature cell as orthorhombic *Pnma*. The charge order that is believed to underly these modulations is thought to result from a strong electron-lattice coupling that localises Mn valence charges in certain (200) planes only. These planes, which have hitherto been understood[1-4] to appear as stripes when viewed in cross-section, have been described in terms of the idealised cations $Mn^{3+}$ and $Mn^{4+}$ despite the high Coulomb cost of complete charge disproportionation. However, various authors have previously cast doubt upon this interpretation[6-9]. In this Letter we show that modulated manganites of the form $La_{1-x}Ca_xMnO_3$ possess a superlattice with uniform periodicity and cannot be described by two Mn charge stripe species.

Polycrystalline samples of $La_{1-x}Ca_xMnO_3$ with grain size ~2 µm were prepared by repeated grinding, pressing and sintering of $La_2O_3$, $CaCO_3$ and $MnO_2$ in stochiometric proportions. The $La_2O_3$ was heated overnight prior to use in order to dehydrate it. Each sample was prepared by initially heating at 950°C for 12 hours to decarboxylate the $CaCO_3$, and then 1350°C for 12 hours. Each sample was then reground, repelleted and heated at 1350°C for 4 days; and then reground, repelleted and reheated at 1300°C for 2 days. X-ray powder diffraction confirmed the presence of a single phase. The macroscopic stoichiometry of our polycrystalline samples is accurate to within 0.1%.

Electron transparency was achieved by conventional mechanical polishing and argon-ion thinning to around 100 nm at liquid nitrogen temperatures.

Electron microscopy data were taken at 90 K over a timescale of seconds using a Philips CM30. Small variations in **q** could be mapped with high precision for reflections with **q** ~ **g-q** (where **g** is a reciprocal lattice vector of the unmodulated structure, e.g. **a***). This is because in dark field images taken using objective apertures that include the two nearby wavevectors, beating creates interference fringes whose spacing $1/|\mathbf{g}-2\mathbf{q}|$ may be measured accurately.

In each grain, **q** was essentially parallel to **a***. Moreover, the wavenumber $q$ in each grain was always found to vary by less than 1% from region to region. For example, using the interference method discussed above to study 80 distinct 50 nm diameter regions in one particular grain of $La_{0.48}Ca_{0.52}MnO_3$, we found $q/a^*=0.446\pm0.004$. No variations in **q** were observed near grain boundaries, except in one specific instance[16]. The absolute value of $q$ varied slightly from grain to grain with a standard deviation of a few percent (indeed the grain used for Fig. 3 has a different value of $q/a^*$ to the one discussed above). However, this intergranular variability has no bearing on our finding that $q$ is highly uniform within each grain. Also, we note that a lock-in transition below our base temperature of 90 K is unlikely. This is because at 90 K, the variation of $q(T)$ is weak (Fig. 1), and large area diffraction patterns yield the expected[1-11] systematic relationship $q/a^* \approx (1-x)$.

Before presenting our key results, let us consider the previously proposed stripe picture[3,4]. The dopings $x=1/2$, $2/3$, $3/4$ correspond to modulations whose periods are integer multiples of the lattice parameter $a$. It is believed[3,4] that only integer period modulations like these are stable. At intermediate (arbitrary) values of $x$ it is believed[3,4] that there are local fluctuations in periodicity via a fine scale "phase separation" into adjacent integer period sub-units, according to the lever rule. The sub-units at $x=2/3$ and $3/4$ are believed[3,4] to contain extra $Mn^{4+}$ planes that constitute charge stacking faults.

In $La_{1-x}Ca_xMnO_3$ at $x=1/2$, the $q/a^*=0.5$ modulation has been interpreted[1-4] in terms of orbitally ordered $Mn^{3+}$ (100) planes alternating with $Mn^{4+}$ (100) planes. If the charge balance is upset slightly by increasing $x$ to 0.52, one would expect a small number of extra $Mn^{4+}$ planes, each corresponding to the presence of one $x=2/3$ sub-unit — Fig. 2(a). A diffraction probe that is significantly smaller than the separation of these $Mn^{4+}$ stacking faults should therefore record $q/a^*=0.5$ most of the time. We stress that the stacking faults, also known as discommensurations, represent sharp discontinuities in charge order. However, in this Letter we demonstrate that no such stacking faults exist and that $q$ is uniform (but not necessarily sinusoidal) — Fig. 2(b).

We now present our key results for $La_{0.48}Ca_{0.52}MnO_3$. The diameter of the diffraction probe is initially above, and then below, the expected stacking fault separation. Fig. 3(a) shows an electron diffraction pattern taken from a 500 nm span within a single grain of $La_{0.48}Ca_{0.52}MnO_3$. As expected, the modulation $q/a^* = 0.468 \pm 0.003$ is close to the expected value of $q/a^* \approx (1-x)=0.48$, c.f. x-ray[5-6,9], neutron[5-6,9,10] and electron diffraction[1-4,7,11] experiments. Fig. 3(b) shows the corresponding measurement when the electron beam is converged to span just 3.6 nm. This probe is smaller than the separation of $Mn^{4+}$ stacking faults which would occur on average every 9.6 nm if one assumes $x=1/2$ and $x=2/3$ sub-units[3,4], or every 6.8 nm if the sub-units are $Mn^{3+}$ and $Mn^{4+}$ planes. Surprisingly, in repeated samplings within this and other grains, we never see the expected $q/a^*=0.5$ that would correspond to orbitally ordered alternating $Mn^{3+}/Mn^{4+}$ planes. Instead, our convergent beam electron diffraction (CBED) probe records $q/a^* = 0.473 \pm 0.005$. Therefore, within experimental error, we find that $q$ does not vary when the diameter of the diffraction probe is changed from above to below the separation of the previously proposed stacking faults. We therefore conclude that the periodicity of the modulation in $La_{0.48}Ca_{0.52}MnO_3$ is uniform. Consequently, we propose that there are no stacking faults.

We reached the same conclusion in an independent study that also involved doping between $x=1/2$ and $x=2/3$. In Fig. 4 we compare the superlattice peaks in large area diffraction patterns with simulations based on the $x=1/2$ and $x=2/3$ sub-units discussed earlier. The simulated superlattice peak when $x=1/2$ was set to match experiment by convoluting a suitable Gaussian with the FFT of a real space waveform built from a run of $x=1/2$ sub-units — Fig. 4(a). This Gaussian was then used in our other simulations, presented in Figs. 4(b-d). At $x=2/3$, simulations based a run of $x=2/3$ sub-units show good agreement with experiment — Fig. 4(d). However, at the intermediate dopings $x=0.52$ and 0.58, simulations based on a random mixture of $x=1/2$ and $x=2/3$ sub-units produce superlattice peaks that are broad with respect to experiment — Figs. 4(b&c). Therefore large area diffraction patterns cannot be reconciled with a picture in which intermediate dopings are represented as random mixtures of end-member sub-units. Moreover, the observed standard deviation in $q/a^*$ of 0.004 in our ~2 μm grains can only be achieved with unrealistically large simulations of ~10 μm. We note that simulations were not sensitive to the amplitude profile in each sub-unit (the two sub-units species used to create Fig. 4 were based on linescans from high resolution electron micrographs[3] adjusted so that the endpoints of each sub-unit matched up). We further note that a quasi-periodic arrangement of sub-units at $x$=0.52 would produce an even broader peak than the random mixture model.

Before interpreting our discovery of uniform periodicity in $La_{1-x}Ca_xMnO_3$, we shall discuss the previous evidence for the stripe picture. This has come primarily from contrast seen in high resolution electron microscopy (HREM) images[2-4]. However, HREM contrast varies strongly as a function of sample thickness, defocus and tilt, and quantitative interpretation is particularly difficult for thicker (50-100 nm) specimens in which dynamical diffraction (multiple scattering) enhances the intensities of the weaker reflections. The other evidence for stripes comes from fringes in dark-field images that have also been interpreted[1] as stacking faults in charge order. However, these fringes arise due to the interference effect explained in paragraph 3, and thus do not[17] represent a real-space phenomenon.

In our electron diffraction experiments we cannot measure directly the nature of the observed modulations. The precise interpretation of our findings is that the modulation has a uniform periodicity but is not necessarily sinusoidal. This discovery of uniform periodicity calls into question the concept of charge order in $La_{1-x}Ca_xMnO_3$. If the superlattice and undistorted parent lattice periodicities are not locked because the electron-lattice coupling is too weak, then the electron-lattice coupling could be too weak to bring about a strong localisation of electronic charge. This would be significant since modulations in manganites have hitherto been associated with charge order and strong electron-lattice coupling.

The weak electron-lattice coupling scenario that we propose would permit intergranular variations in strain to generate the observed intergranular variability in $q/a^*$ of <1% (see paragraph 4). However, intergranular variability due to strain can also be reconciled with the previously proposed charge stacking faults[1-4] because the local elastic properties should be correlated with the local arrangement of cations such as $Mn^{3+}$ and $Mn^{4+}$. Therefore intergranular variability in $q/a^*$ does not represent evidence for weak coupling. However, according to simulations that we have performed based[18] on the Frenkel Kontorova model, the intragranular variability in $q/a^*$ of <1% places an upper bound of 3% on a parameter that may be crudely compared with the electron-lattice coupling constant.

If there is weak electron-lattice coupling and therefore some degree of electron itineracy, one might expect metallic behaviour above the long-range ordering temperature. We argue that this is not seen[19] due to short-range fluctuations. For example, we have observed diffuse superlattice spots in $La_{0.48}Ca_{0.52}MnO_3$ at room temperature (reminiscent of those seen[20] at $x$=1/3). Moreover, an optical "pseudogap" is known to persist to even higher temperatures[19]. We note that structural modulations in layered manganites[21-25] are also associated with some degree of electron itineracy, having been attributed to enhanced Fermi surface nesting due to reduced dimensionality.

Uniform periodicity necessarily rules out the stripe picture. This is because it is impossible to reconcile uniform periodicity with two Mn charge stripes as described by the idealised cations $Mn^{3+}$ and $Mn^{4+}$; even a primarily structural modulation along **q** would have implications for the Mn valence states. One observation consistent with multiple Mn valence states is that **q** and **a\*** are not always precisely co-linear (this work and [7,11]).

Our results do not rule out the possibility of complex 3D distributions of say the idealised cations $Mn^{3+}$ and $Mn^{4+}$, although superstructure is observed only in the **a\*** direction [11]. A modulated variation of the $Mn^{3+}$:$Mn^{4+}$ ratio in successive planes perpendicular to **a\***, without in-plane cation ordering, could account for the uniform 1D periodicity. However, the disorder of cations within the planes would tend to broaden the diffraction peaks, whereas sharp spots are observed. The proposed scenario of weak electron-lattice coupling is the simplest explanation consistent with our data.

Prior to publication, our findings were presented to Brey and they inspired him to study[26] the ground state of $La_{1-x}Ca_xMnO_3$ by self-consistently solving a realistic Hamiltonian with parameters that correctly reproduce the ferromagnetic phase at $x$=0.3 and the CE antiferromagnetic phase at $x=0.5$. Brey finds for $x=0.5$ that a weak uniform modulation of the charge is stable with respect to the previously proposed charge ordered phases, and the "phase separation" discussed earlier. Furthermore, Brey finds that uniform periodicity is even possible in the limit of zero electron-lattice coupling (due to orbital order), and this supports the interpretation proposed here.

We note that there is now both experimental and theoretical evidence to suggest that Coulombically expensive variations in valence are small in the modulated manganites. For example, resonant x-ray scattering[8] has previously been used to suggest this, a full neutron refinement[27] of the half-doped manganite $Pr_{0.5}Ca_{0.5}MnO_3$, records a Mn ion charge difference that corresponds to only 25% of the difference between the ideal $Mn^{3+}$ and $Mn^{4+}$ cations, and neutron pair distribution functions for $La_{0.5}Ca_{0.5}MnO_3$ are consistent with the non-integral Mn valence states that we suggest here based on our experiments at higher $x$[28]. Moreover, recent *ab initio* Hartree-Fock calculations[29] at $x=1/2$, and the solutions of Brey described above, support this view.

Whatever the nature of the modulated phases of $La_{1-x}Ca_xMnO_3$, the absence of a lock-in to integer period modulations in $x>0.5$ is surprising, particularly[30] near $x=0.5$. A lock-in to $q/a^*=0.5$ is however seen[31] in modulated manganites with $x<0.5$. This could be due to coexisting ferromagnetic order (which at $x=0.5$ causes deviation from $q/a^*=0.5$, rather than a lock-in, at sufficiently high temperatures)[32]. Indeed, as previously discussed, the coexistence of ferromagnetism and charge order[33] can lead to complex phase structures[34]. Above $x=0.5$ it would appear that there is no physical significance associated with integer period modulations that correspond to dopings of the form $x=2/3$ and $3/4$. As argued earlier, modulations in $La_{1-x}Ca_xMnO_3$ cannot be described as stripes of charge order. Instead, we suggest that these modulated manganites possess small amplitude[25-29] charge density waves with non-zero and possibly high electron itineracy. It would be interesting to see whether manganite compositions with smaller A-site cations also fail to produce the anticipated lock-in behaviours. Our findings demonstrate that manganite physics is becoming ever more subtle.

We thank L. Brey for valuable comments, and also E. Artacho, M. J. Calderón, S. N. Coppersmith, M. T. Fernández-Díaz, V. Ferrari, L. E. Hueso, Š. Kos, J. L. Martínez, A. J. Millis, G. C. Milward, P. G. Radaelli, C. N. R. Rao, E. J. Rosten and J. M. Tranquada. This work was funded by the UK EPSRC and the Royal Society.

FIG. 1. Temperature variation of $q$ in one grain of $La_{0.48}Ca_{0.52}MnO_3$. The low temperature value does not match the value of $q/a^*=1-x=0.48$ associated with the nominal composition. This could be due to some combination of thermal history, strain, surfaces and even intrinsic effects. The weak variation of $q(T)$ at 90 K suggests that there is no reason to expect a lock-in transition below our base temperature of 90 K.

FIG. 2. 1-D representation of $La_{0.48}Ca_{0.52}MnO_3$. The Mn atoms in the (200) planes of the parent lattice are represented as a linear chain of dots separated by $a/2$. (a) In the conventional charge order picture[3,4], orbitally ordered $Mn^{3+}$ planes alternate with $Mn^{4+}$ planes, and charge neutrality is maintained with an extra $Mn^{4+}$ plane on average every 9.6 nm assuming $x=1/2$ and $x=2/3$ sub-units[3,4], or every 6.8 nm if the sub-units are $Mn^{3+}$ and $Mn^{4+}$ planes. (b) In this Letter we show that the superlattice modulation is uniform and that there are no charge stacking faults. Note that we have not resolved the nature of the modulation in our diffraction experiments, and cannot confirm whether it is sinusoidal or not.

FIG. 3. (a) Electron diffraction pattern from a $La_{0.48}Ca_{0.52}MnO_3$ grain at 90 K taken with a selected area aperture of diameter 500 nm (as measured in the object plane). Parent reflections are indicated by solid lines. The wavenumber of the modulation $q/a^* = 0.468 \pm 0.003$. (b) Convergent beam electron diffraction (CBED) pattern from the same grain. The superlattice period $2p/q \sim a/(1-x) = 1.1$ nm < CBED probe FWHM (3.6 nm) < the expected $Mn^{4+}$ stacking fault separations of 9.6 nm or 6.8 nm discussed in the text. In all CBED patterns acquired within this and other grains, we never see the expected[3,4] $q/a^*=0.5$ between stacking faults (indicated by arrows and dashed lines). Instead, the modulation is uniform and $q/a^* = 0.473 \pm 0.005$. This matches the large area value within experimental error.

FIG. 4. Electron diffraction patterns (least saturated areas) and the corresponding linescans (solid lines) for $La_{1-x}Ca_xMnO_3$ with (a) $x=0.5$, (b) 0.52, (c) 0.58 and (d) 0.67 taken with a selected area aperture of diameter 200 nm at 90 K looking down the [010] zone axis. The superlattice peaks are always sharp. By contrast, simulations (dotted lines) that are based on $x=1/2$ and $x=2/3$ type sub-units[3,4] show broad superlattice peaks when there is a random mixture at intermediate dopings (b&c). Note that absolute values of $q$ are subject to the details of the model, and that extrinsic effects including multiple scattering enhance the second superlattice peak in (b-d) and the parent lattice peak in (a-d).

Figure 1

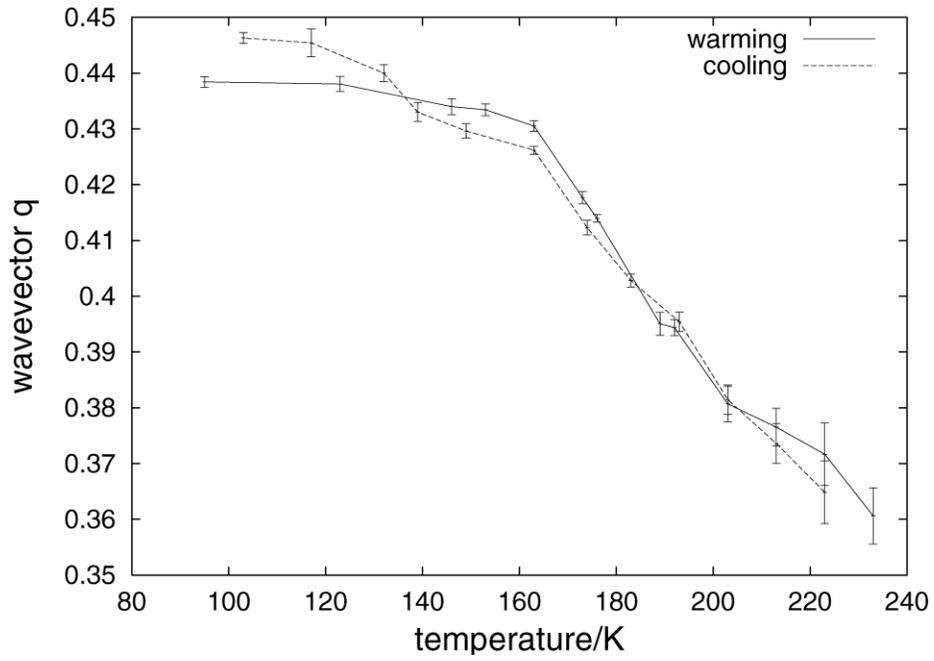

Figure 2

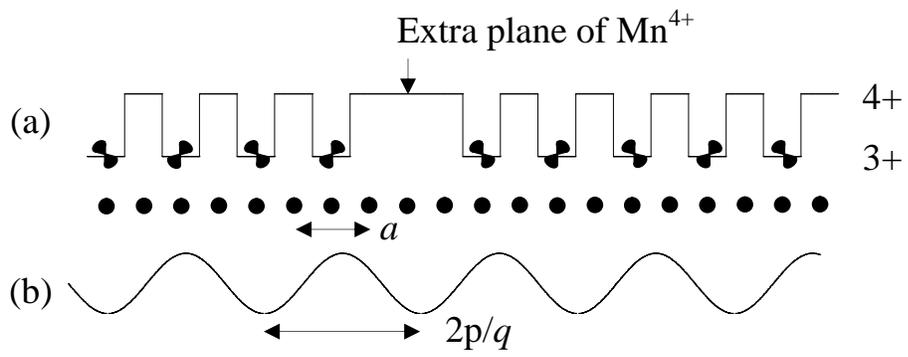

Figure 3

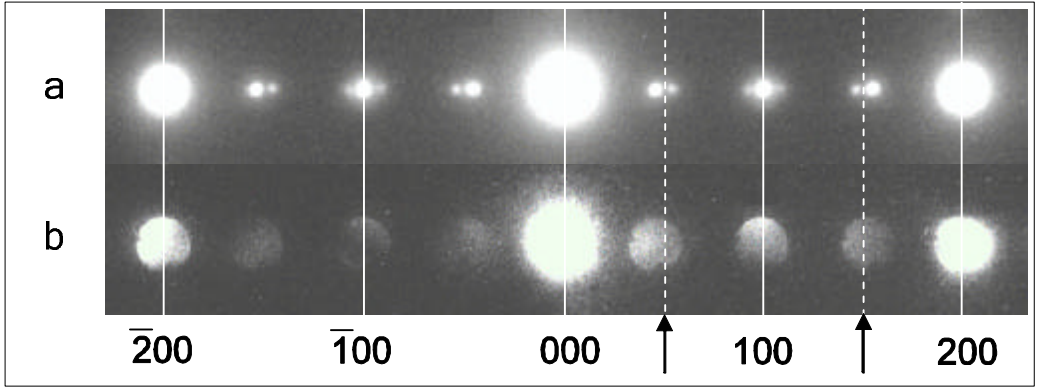

Figure 4

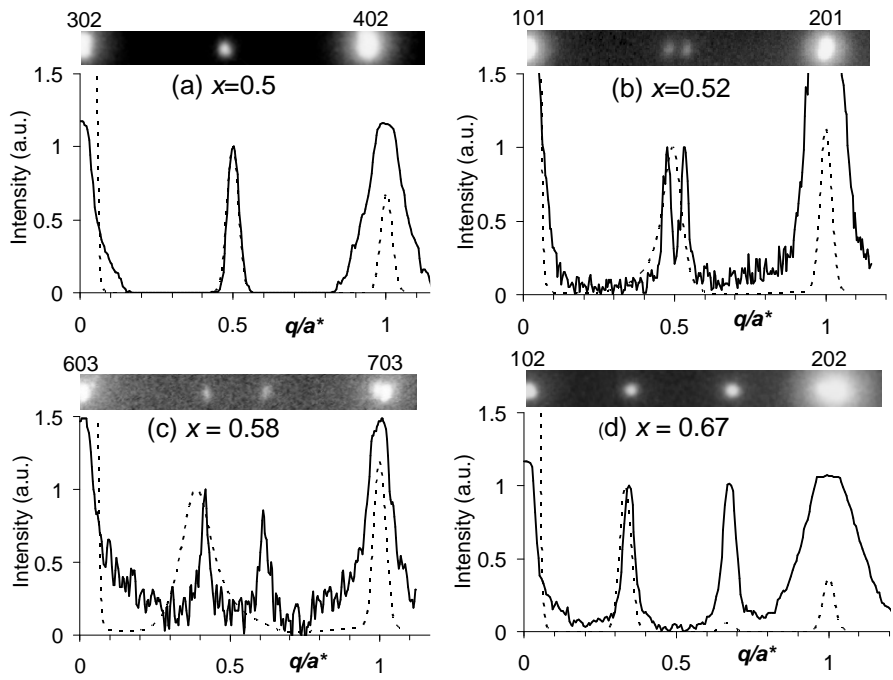